\pdfoutput=1
\documentclass[aps, prl, reprint,nofootinbib]{revtex4-1}
\usepackage{amsmath, amsfonts, amssymb, hyperref, color, graphicx, xspace, enumitem, slashed}
\usepackage[T1]{fontenc}
\definecolor{nicered}{rgb}{.7,.1,.1}
\definecolor{nicegreen}{rgb}{.1,.5,.1}
\definecolor{darkblue}{rgb}{0,0,.5}
\hypersetup{colorlinks, citecolor=darkblue, linkcolor=darkblue, urlcolor=darkblue}

\usepackage{tikz}
\usetikzlibrary{arrows,shapes}
\usetikzlibrary{trees}
\usetikzlibrary{matrix,arrows} 				
\usetikzlibrary{calc} 
\usetikzlibrary{positioning}					
\usetikzlibrary{calc,through}				
\usetikzlibrary{decorations.pathreplacing}  	
\usepackage[tikz]{bclogo} 					
\usepackage{pgffor}						

\usetikzlibrary{decorations.pathmorphing}		
\usetikzlibrary{decorations.markings}
\usetikzlibrary{intersections}				

\newcommand{\be}{\begin{equation}}
\newcommand{\ee}{\end{equation}}

\usetikzlibrary{decorations.pathmorphing}	
\usetikzlibrary{decorations.markings}
\usetikzlibrary{intersections}				
\tikzset{
    vector/.style={decorate, decoration={snake}, draw},
    graviton/.style={decorate, decoration={snake,amplitude=1.5pt}, draw},
    fermion/.style={postaction={decorate},
        decoration={markings,mark=at position .55 with {\arrow{>}}}},
    fermionbar/.style={draw, postaction={decorate},
        decoration={markings,mark=at position .55 with {\arrow{<}}}},
    fermionnoarrow/.style={},
    gluon/.style={decorate,
        decoration={coil,amplitude=4pt, segment length=5pt}},
    scalar/.style={dashed, postaction={decorate},
        decoration={markings,mark=at position .55 with {\arrow{>}}}},
    scalarbar/.style={dashed, postaction={decorate},
        decoration={markings,mark=at position .55 with {\arrow{<}}}},
    scalarnoarrow/.style={dashed,draw},
	provector/.style={decorate, decoration={snake,amplitude=2.5pt}, draw},
	antivector/.style={decorate, decoration={snake,amplitude=-2.5pt}, draw},
	    electron/.style={draw=black, postaction={decorate},
        decoration={markings,mark=at position .55 with {\arrow[draw=black]{>}}}},
	bigvector/.style={decorate, decoration={snake,amplitude=4pt}, draw},
	vectorscalar/.style={loosely dotted,draw=black, postaction={decorate}},
}

\usepackage{slashed}

\begin{document}

\setcounter{secnumdepth}{2}

\title{\LARGE Supercooling exit from charge supersaturation}

\author{Pietro Baratella}
\email{pietro.baratella@ijs.si}
\affiliation{Jo\v{z}ef Stefan Institute, Jamova 39, 1000 Ljubljana, Slovenia}

\begin{abstract}
\noindent Systems that feature a scalar field $\phi$ with a quasi scale invariant potential, metastable at $\phi=0$, can remain trapped, during cosmic evolution, in the `wrong' vacuum because the process of bubble nucleation to the true vacuum is inefficient. If $\phi$ carries a conserved charge $Q$, the presence in the system of a non-zero chemical potential for $Q$ offers the possibility of escaping eternal supercooling: when a species decouples from the plasma the balance among the stabilising effect of temperature and the destabilising effect of chemical potential can change in favour of the latter, so that $\phi$ condenses and triggers the transition to the stable phase.
\end{abstract}

\maketitle

\section{Introduction}

Scalar fields with a quasi scale invariant potential that is metastable at the origin appear in various problems. To mention a few: old models of Inflation \cite{Guth:1980zm}, the Coleman-Weinberg model \cite{PhysRevD.7.1888}, ultraviolet completions of the electroweak sector with strong dynamics \cite{Creminelli:2001th}, and last but not least the Standard Model at regimes in which $v_{\rm EW}$ can be neglected \cite{Degrassi:2012ry,Buttazzo:2013uya,Baratella:2024hju}. 

Potentials of this kind can be problematic in a cosmological context. At very early times, the temperature is so high that the potential is generally stabilised at $\phi=0$ due to a thermal contribution of the form $\Delta V=aT^2\phi^2$, with $a>0$. As the universe cools down a local minimum forms at some $v(T)\neq 0$, and the new minimum becomes degenerate with the symmetric one for some critical temperature $T_c$. Further cooling makes the symmetry breaking minimum energetically favoured. 

However, due to the flatness of the zero temperature contribution to the potential, thermal effects leave a metastable minimum at $\phi=0$ that persists at arbitrarily low temperatures, until the cosmic plasma reaches the de Sitter temperature $T_{\rm dS}$ where no further cooling is possible.

If no drastic departure from this picture takes place, the only way in which a phase transition to the stable phase can take place is via bubble nucleation. This process is exponentially suppressed and is not necessarily efficient enough. For a quasi scale invariant potential $V=-\lambda(\phi)\phi^4$, the tunnelling rate $\gamma$ is proportional to $e^{-2\pi^2/3\lambda}$ and therefore approximately constant for $T_{\rm dS}<T<T_c$. If $\lambda$ is small then $\gamma\ll H^4$, where $H$ is the Hubble parameter, and the transition to the true vacuum cannot take place. Because of the cooling induced by the universe expansion, the vacuum energy soon starts dominating the energy density of the system and the universe starts inflating; $H$ becomes constant and, since $\gamma$ is also approximately constant, nucleation can never become efficient.

For most of the previously mentioned scenarios, it is phenomenologically necessary to eventually transition to the stable vacuum. The exception is the Standard Model, where the false vacuum must survive cosmic evolution.

In the seminal paper \cite{Witten:1980ez}, in the context of a Coleman-Weinberg implementation of the Higgs sector, it was noted that supercooling proceeds all the way to $T_{\rm QCD}$, at which temperature the condensation of quarks induces a linear term to the Higgs potential, of the form $\Delta V=y  \phi\langle \bar q q\rangle$. This term -- which comes from the Yukawa coupling of the Higgs to the quarks -- destabilises the symmetric point $\phi=0$ and drives the transition to the true vacuum, which happens via classical roll-down instead of quantum or thermally assisted tunnelling.

\

The present work proposes an alternative mechanism for exiting eternal supercooling, based on the well known fact that a chemical potential for $\phi$ gives a destabilising contribution to the effective potential, of the form $\Delta V=-\mu^2|\phi|^2$, where $\mu$ is the chemical potential for $\phi$. When $\mu>\sqrt a T$ the point $\phi=0$ turns from a local minimum to a maximum and the field rolls down.

The example of \cite{Witten:1980ez} shows that, if no physical scale other than temperature breaks the (approximate) scale invariance of the problem, supercooling does not end. On the contrary, the existence of the QCD scale $\Lambda_{\rm QCD}$ and a coupling between the QCD sector and the Higgs is at the heart of the mechanism for supercooling exit (see also \cite{Hambye:2018qjv} and \cite{Baratella:2018pxi} for more recent investigations in the context of Coleman-Weinberg-like models and theories with a strongly coupled electroweak sector, respectively).

In the examples that are going to be presented here the same is observed. Suppose that, at some temperature $T<T_c$, some charge excess is present that makes $\mu\neq 0$. For $\phi=0$ to be initially an equilibrium point, it must be $\mu<\sqrt a T$. If the particles that make up the thermal plasma are all massless or have negligible masses with respect to $T$, the equilibrium evolution of the system implies that the ratio $\mu/T$ remains constant. Therefore the coefficient in front of $|\phi|^2$ remains positive and the presence of $\mu$ does not change the above qualitative picture.

The situation changes drastically whenever the systems cools down to $T\lesssim \Lambda$, with $\Lambda$ some physical scale that characterises the plasma (typically the mass of a particle, dynamically generated or not). In this case, the ratio $\mu/T$ in general changes and the condition $\mu<\sqrt a T$ can be violated, making the system unstable.

The remainder of the paper is organised as follows: in Section~\ref{sec:discussion} the stage is set, and the condition for inducing Bose-Einstein condensation for $\phi$ (which eventually triggers the end of supercooling) is derived; Section~\ref{sec:mass} presents formulas for the thermal mass of $\phi$ at leading order in the couplings that govern its interaction with the other particles of the thermal plasma; Section~\ref{sec:examples} considers several examples of the mechanism and Section~\ref{sec:conclusions} draws general conclusions.

\begin{figure}
  \vspace{.1cm}
  \centering
  \includegraphics[width=.8\columnwidth]{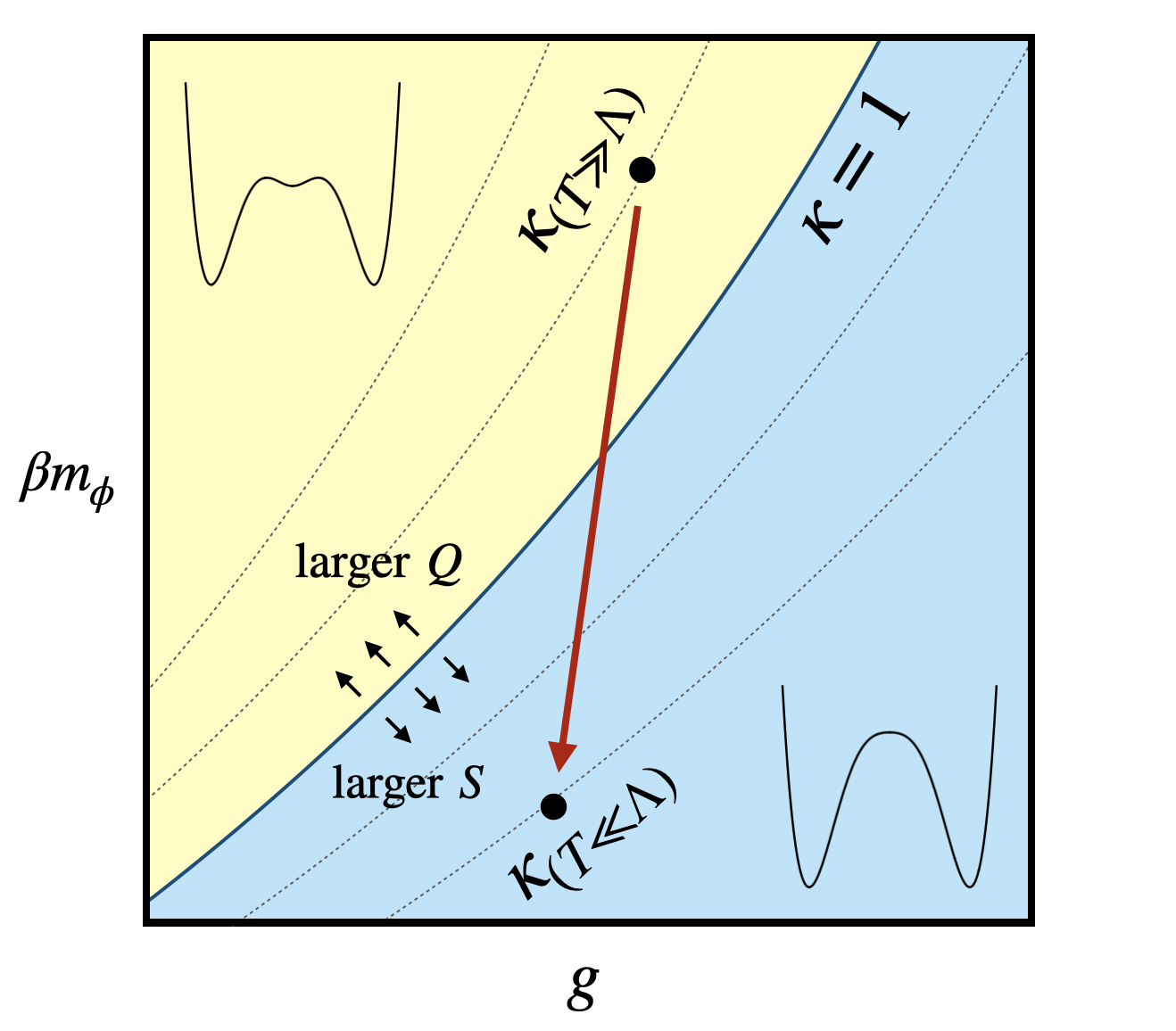}
  \caption{Contours of $\kappa$, defined in (\ref{master}), at fixed entropy $S$ and charge $Q$, as a function of the number of active species $g$ and the ratio of the scalar's thermal mass to temperature $\beta m_\phi=\sqrt a$ (schematic). Both $g$ and $\sqrt a$ are approximately piecewise constant functions of $T$ that decrease under cooling when $T$ crosses a physical scale $\Lambda$. The change in these two parameters can be such that the system goes from the yellow region ($\kappa<1$) to the blue region ($\kappa>1$) where a condensate is formed. The contours of $\kappa$ can also be thought of as different critical lines ($\kappa=0$) at varying $Q$ or $S$.}
  \label{fig:param}
\end{figure}

\section{Mechanism}\label{sec:discussion}

We consider a plasma of particles that are in thermal equilibrium at temperature $T$. Some of the species, including $\phi$, are charged under a global conserved charge $Q$. We are going to study regimes in which particles have either $m\gg T$, and so their density is Boltzmann suppressed, or $m\ll T$, so they can be treated as relativistic species.

The presence of a charge unbalance implies that the system has a chemical potential $\mu$ different than zero. For bosonic species, a large enough chemical potential induces Bose-Einstein condensation. Quantitatively, the condition is that
\be\label{mu}
\mu>m_\phi\,,
\ee
where $m_\phi$ includes thermal corrections that typically dominate at high temperatures.

The simple condition for condensation in (\ref{mu}) gives a concrete meaning to $\mu$, whose value is driven by the cosmic evolution. The relevant questions are then: What is the path that $\mu$ follows in the cosmic evolution? Is it possible to reach the condition (\ref{mu}) if it is initially not satisfied?

To address these questions some assumptions have to be made. It is physically natural and mathematically elegant to assume that (i) the system experiences no drastic departure from equilibrium, so its entropy $S$ does not increase, and (ii) that $Q$ is exactly conserved.

$S$ and $Q$ are uniquely determined by $T,\mu$ and $V$, with a trivial dependence on the volume ($S,Q\propto V$). Their ratio is both conserved and independent of volume, so one gets an expression like
\be\label{sigma}
\frac SQ=f(T,\mu)\,,
\ee
which can be inverted to determine $\mu(T)|_{S/Q}$. This gives the chemical potential for $\phi$ as a function of temperature at fixed entropy per charge. 

To derive an expression for $\mu(T)|_{S/Q}$ one needs to determine the function $f$ in (\ref{sigma}). Let us start from estimating the entropy. This is largely dominated by the relativistic species and is essentially insensitive to $\mu$ if $\mu\lesssim T$.\footnote{In the examples that we are going to consider in Section~\ref{sec:examples}, this condition is always satisfied at weak coupling.} The free theory expression at vanishing $\mu$, {\it i.e.}
\be\label{entropy}
\frac SV=\frac{2\pi^2T^3}{45}\big(N_b+\tfrac78N_f\big)= gT^3\,,
\ee 
where $N_b$ and $N_f$ are respectively the number of relativistic bosons and fermions, is enough to capture it.

Assuming for simplicity that the only particle which is charged under $Q$ is $\phi$ itself, and that it has charge $q_\phi$, one has
\be\label{charge}
\frac Q V=q_\phi (n_\phi- n_{\bar\phi})\,,
\ee
where the number density of particle $\phi$ is given by
\be\label{Ndensity}
n_\phi=\int \frac{{\rm d}^3k}{(2\pi)^3}\frac1{e^{\beta(\sqrt{k^2+m_\phi^2}-q_\phi\mu)}- 1}\,,
\ee
while $n_{\bar\phi}$ is obtained by changing $q_\phi\to q_{\bar\phi}=-q_\phi$.

Unlike with the entropy, the role of $\mu$ is crucial for determining the total charge $Q$. In particular, $Q=0$ when $\mu=0$ because the contributions of particle and antiparticle exactly cancel (for the entropy, instead, particle and antiparticle contribute with the same sign). With a potential for $\phi$ which is approximately scale invariant at zero temperature, the mass of $\phi$ comes mostly from thermal effects, and is given by $m_\phi=\sqrt a T$, with $a$ some expression of the marginal couplings.\footnote{The way $m_\phi^2$ enters in (\ref{Ndensity}) implies a resummation of the so-called `daisy' diagrams. See for example \cite{Arnold:1994ps}; see also \cite{bar} for a recent analysis on the emergence of a thermal mass in purely $S$-matricial terms.} By simple dimensional analysis one can express $n_\phi=T^3 I(\tfrac\mu T)$, with
\be\label{I}
I(x)=\frac1{2\pi^2}\int_0^\infty {\rm d} t\, t^2 \frac1{e^{\sqrt{t^2+a}-q_\phi x}-1}\,.
\ee
The expression in (\ref{sigma}) can then be refined by stating that entropy per charge is a function of only \emph{the ratio} of chemical potential and temperature. Precisely
\be
\frac Q S=g^{-1}q_\phi \big({I(\tfrac\mu T)-I(-\tfrac\mu T)}\big)\,.
\ee
Inverting the expression one finds
\be\label{nu}
\mu=\alpha\big[\frac S{gQ},a\big] \,T\,,
\ee
where $S/Q$ is constant in time if there is no charge nor entropy injection, and similarly $a$ and $g$ unless $T$ crosses some physical mass scale (the form of $\alpha$ is not important for the moment).

The condition that defines the onset of Bose-Einstein condensation, {\it i.e.} $q_\phi \mu>\sqrt a T$, translates into
\be
q_\phi\alpha>\sqrt a\,,
\ee
which depends on temperature only through $g$ and $a$.

The question now becomes the following: is it possible for $g$ and $a$ to change in such a way that condensation is induced when going from higher to lower temperatures?

The physics that governs changes in $g$ is pretty transparent, and makes us lean towards a negative answer to the question. This is because $g$ decreases under cooling when particles become non-relativistic and their density gets Boltzmann suppressed. A decrease of $g$, in turn, is equivalent to an increase of entropy per unit charge at $g$ fixed, as can be seen from (\ref{nu}). With more and more entropy per unit charge, the system gets further and further away from criticality (see also (\ref{master}) and the discussion below) and condensation cannot take place upon cooling unless a compensating effect comes from the change in $a$.

Before going into a discussion of the physics of $m_\phi$ and a quantitative determination of $a$ in Section~\ref{sec:mass}, it is useful to derive an approximate formula for $I(\nu)$ that makes the dependence on the various parameters explicit.

\subsection*{Simple parametric expressions}

A useful approximation to $n_\phi-n_{\bar\phi}$ is derived by Taylor expanding (\ref{I}) and discarding $O(\nu^3)$ terms. The leading order cancels in the difference $I(\nu)-I(-\nu)$, while the coefficient of the linear term is a function of $a$ with a finite $\lim_{a\to0}$. In many scenarios $a$ is a small number and it is a good approximation to just take the limit. Doing the integral one finds simply
\be
\frac QV=\frac{q_\phi^2}3 \,\mu\, T^2\,.
\ee
After taking the ratio with (\ref{entropy}) one finds $S q_\phi^2\mu=3gQT$, which can be used to derive the following expression for the chemical potential:
\be\label{mu_sigma}
\mu=\frac{3gQ}{S q_\phi^2}\,T\,.
\ee
Condensation takes place when $q_\phi\mu>\sqrt a T$, that is when
\be\label{master}
\frac{3gQ}{q_\phi\sqrt a S}\equiv\kappa>1\,.
\ee
This expression can be read in many ways. Perhaps most intuitively as a condition on the maximal charge that can be stored in the system before inducing condensation (similarly to supersaturation effects in chemistry). Thinking at $Q/S$ fixed, as appropriate for cosmic evolution, one concludes that the condition for $\kappa$ to grow under cooling is that the relative decrease in $\sqrt a$ is stronger than the relative decrease of $g$. This is illustrated in Figure~\ref{fig:param}. As we are going to see in Section~\ref{sec:mass}, this means that particles that Boltzmann decouple from the plasma are the ones that give a sizeable contribution to the thermal mass of $\phi$.

In the limit of small $a$, it is immediate to derive an approximate expression for $Q$ when particles other than $\phi$ are charged
\be\label{charge2}
\frac QV =\frac13\mu T^2\bigg(\sum_b q_b^2+\frac12 \sum_f q_f^2\bigg)\equiv \frac12\mu T^2{\cal Q}^2\,,
\ee
valid up to $O\big((\beta\mu)^2\big)$ corrections, with $b$ a bosonic and $f$ a fermionic species. A decrease in the number of \emph{charged} relativistic species tends to increase $\mu/T$, as one can see by promoting $q_\phi^2\to {\cal Q}^2$ in  (\ref{mu_sigma}). Condensation happens when $\kappa=3gq_\phi Q/({\cal Q}^2\sqrt a S)>1$. When several scalars $i$ are present, one has a different $\kappa_i$ factor for each of them, with
\be
\kappa_i=\bigg(\frac{3gQ}{{\cal Q}^2S}\bigg) \frac{q_i}{\sqrt{a_i}}\,.
\ee
Particles with the largest ratio ${q_i}/{\sqrt{a_i}}$ are more prone to condensation.\

\section{Thermal mass}\label{sec:mass}
Given that $\phi$ has vanishing or negligible mass at zero temperature, its mass at finite temperature is entirely given by thermal effects and is set by $T$. In these conditions, at leading order in couplings the thermal mass squared of $\phi$ is captured by the one-loop self-energy at vanishing momentum, {\it i.e.} $m_\phi^2=-\lim_{{\bf p}\to 0}\Sigma^{(1)}_T(0,\bf{p})$, where the suffix $T$ restricts to the thermal part of the self-energy \cite{Kapusta:2006pm}.

$\Sigma_T^{(1)}$ can be interpreted as coming from the forward scattering (at tree level) of $\phi$ with the particles that constitute the thermal bath \cite{Schenk:1993ru}. The contribution of a given mode of species $i$ that has momentum ${\bf k}$ is proportional to its density $n_i({\bf k})$. In formulas,
\be\label{mass}
m_\phi^2=\lim_{{\bf p}\to 0}\sum_i \int\frac{{\rm d}^3k}{(2\pi)^32E_i({\bf k})} M_{i\phi\to i\phi}({\bf k},{\bf p}) n_i({\bf k})
\ee
where $M_{i\phi\to i\phi}$ is computed in the forward configuration, with $p_\mu=(0,{\bf p})$ and $k_\mu=(E_i({\bf k}),{\bf k})$, with $E_i({\bf k})=\sqrt{{\bf k}^2+m_i^2}$. It is not always safe to simply set ${\bf p}=0$ in the integrand \cite{Arnold:1992qy}.

The most important feature of (\ref{mass}) is that the contribution of a given particle $i$ is exponentially suppressed as $e^{-m_i/T}$ when $T\lesssim m_i$, due to the presence of $n_i({\bf k})=(e^{\beta E_i({\bf k})}\pm 1)^{-1}$. Given that particles and antiparticles contribute with the same sign, the effect of a chemical potential is not important in (\ref{mass}).

The effect of the interaction of $\phi$ with scalars, spin 1/2 fermions and gauge bosons via the usual marginal couplings is going to be presented next.

\subsection{Scalar quartic}

The contribution to the thermal mass of $\phi$ coming from a quartic coupling ${\cal L}=-\frac12\lambda'|\phi|^2\varphi^2$ is the easiest to compute, as the forward amplitude $\phi\varphi\to\phi\varphi$ is simply given by $\lambda'$. If the scalar $\varphi$ has mass $m_\varphi$, one finds \cite{Laine:2016hma}
\be\label{drop}
m_\phi^2=
\begin{cases} 
       \frac{\lambda'T^2}{24}  & T\gg m_\varphi\\
       \frac{\lambda'}2(2\pi)^{-\frac32}\sqrt{m_\varphi T^3}\,e^{-m_\varphi/T} & T\ll m_\varphi
   \end{cases}\,.
\ee
Given the exponential damping, the contribution of $\varphi$ to the thermal mass of $\phi$ effectively disappears for $T\ll m_\varphi$ and any other contribution quickly dominates over it, be it the zero temperature contribution (assumed here to be negligible for a wide range of temperatures) or the thermal mass coming from those particles of the thermal bath that are still relativistic.

Keeping the same normalisation for the $2\to2$ amplitude, one gets an additional factor of 2 if $\varphi$ is complex and a factor of $d_\varphi$ if it constitutes a multiplet with $d_\varphi$ components.
Notice that a complex $\varphi$ can be either charged or neutral under $Q$, the conserved operator under which we assume $\phi$ to be charged.

\subsection{Yukawa}

Since $\phi$ is charged under $Q$, a Yukawa term has to involve two fermionic species, $\psi$ and $\chi$, with interaction Lagrangian
\be\label{Yukawa}
{\cal L}=\phi \bar\psi\big(y+y'\gamma_5\big)\chi+c.c.
\ee
and masses $m_\psi$ and $m_\chi$ (we are assuming for simplicity that $\psi$ and $\chi$ are Dirac fermions that admit a Dirac mass term). The charges respect $q_\phi+q_\chi=q_\psi$, so at least one fermionic species is charged under $Q$.

\begin{figure}
\vspace{.1cm}
  \centering
      \begin{tikzpicture}[line width=1.2 pt, scale=1, baseline=(current bounding box.center)]

		\draw[fermionbar, rotate=-55] (0,0) circle (1);
		\draw[dotted] (-1,0) -- (-2.45,0);
		\draw[dotted] (1,0) -- (2.45,0);
		\node at (0,-.65) {$\psi$};
		\node at (0,.65) {$\chi$};
		\node at (-2,0.3) {$\phi$};
	
 \end{tikzpicture}
  \caption{The one-loop self energy of $\phi$ in the model specified by (\ref{Yukawa}) comes from the above diagram. Including thermal effects amounts to compute it with thermal instead of vacuum propagators for $\psi$ and $\chi$.}
  \label{fig:loop}
\end{figure}
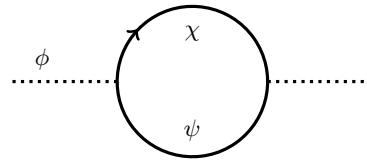

The thermal contribution to the mass of $\phi$ induced by (\ref{Yukawa}) can be computed with (\ref{mass}). With this method one needs to be very careful in keeping into account the relative signs of contributions coming from four species in the plasma, that is $\psi,\bar\psi,\chi$ and $\bar\chi$. Another possibility is to derive all four contributions from one single object: the thermal loop of Figure~\ref{fig:loop}.

It can be computed with the usual (Euclidean) Feynman rules, albeit with a modified definition of the loop integration: instead of an integral over $k_0$, the time component of the loop momentum, one needs to perform a sum over fermionic Matsubara modes $k_0=(2n+1)\pi T$, with $n$ integer.

Setting the external momentum to zero, one has
\be
\Sigma^{(1)}=\int_k\frac{{\rm tr}[(\slashed{k}+m_\psi)(y+y'\gamma_5)(\slashed{k}+m_\chi)(y-y'\gamma_5)] }{(k^2+m_\psi^2)(k^2+m_\chi^2)}
\ee
where $\int_k\equiv T\sum_{k_0}\int\frac{{\rm d}^3k}{(2\pi)^3}$, and $\Sigma^{(1)}$ contains also the (not yet separated) zero temperature contribution. Evaluating the trace one gets
\be
{\rm tr}[\ldots]=4\big(k^2(y^2+y'^2)-m_\psi m_\chi(y^2-y'^2)\big)\,.
\ee
With a standard reduction procedure one can rewrite the squared momentum in the numerator as $k^2=\frac12(k^2+m_\psi^2)+\frac12(k^2+m_\chi^2)$ and reduce the problem to the evaluation of the following thermal integrals
\begin{align}
I_1&=\int_k \frac 1{k^2+m^2} \,,\\
I_2&=\int_k \frac1{(k^2+m^2)(k^2+\hat m^2)}\,.
\end{align}
Let us review the evaluation of the tadpole integral in detail, to see how the separation of zero temperature effects arises algebraically and the on-shell nature of the remaining, purely thermal, contribution.

The key identity to invoke is a version of the Poisson summation formula, which allows to rewrite the sum over Matsubara modes as
\begin{align}
T\sum_{k_0} \frac1{k_0^2+E_{\bf k}^2}&=\sum_{\ell\in {\bf Z}}(-1)^\ell\int_{-\infty}^{+\infty}\frac{{\rm d}\omega}{2\pi}\frac{e^{-i\ell\omega\beta}}{\omega^2+E_{\bf k}^2} \nonumber \\
&=\frac1{E_{\bf k}}\bigg(\frac12+\sum_{\ell>0}(-1)^\ell e^{-\ell\beta E_{\bf k}}\bigg)
\end{align}
with $E_{\bf k}=\sqrt{{\bf k}^2+m^2}$. To go to the second line one needs to interpret the integral in ${\rm d}\omega$ as a contour integral and close the contour in the upper or lower half plane for respectively negative and positive $\ell$. The integral then localises on the poles $\omega=\pm i E_{\bf k}$ by virtue of the Cauchy theorem. The sum over $\ell>0$ can be recognised as \emph{minus} the Fermi-Dirac density $n_f({\bf k})$, while the $\ell=0$ term is the contribution of vacuum fluctuations that one computes with the usual zero temperature Feynman rules.

All in all one finds
\be
I_1=\int\frac{{\rm d}^3k}{(2\pi)^3E_{\bf k}}\bigg(\frac12-n_f({\bf k})\bigg)\,.
\ee
With analogous manipulations one finds, for the $T$-dependent part of $I_2$,
\be
I_2\big|_T=\frac{1}{m^2-{\hat m}^2}\int\frac{{\rm d}^3k}{(2\pi)^3}\bigg(\frac{n_f({\bf k})}{E_{\bf k}}-\frac{\hat n_f({\bf k})}{\hat E_{\bf k}}\bigg)\,.
\ee
In the evaluation of $I_2$ one encounters poles in $\omega=\pm i E_{\bf k}$ and $\omega=\pm i \hat E_{\bf k}$. For a discussion on the subtleties of the computation of $\Sigma^{(1)}$ for coincident masses, see \cite{Arnold:1992qy}. Putting everything together, one finds that the one-loop thermal mass of $\phi$ induced by (\ref{Yukawa}) is given by
\begin{align}\label{ymass}
m_\phi^2=4m_\psi \bigg(\frac{y^2}{m_\psi{-}m_\chi}+\frac{y'^2}{m_\psi{+}m_\chi} \bigg)\int_{\bf k}\frac{n_\psi}{E_\psi}+(\psi\leftrightarrow\chi)
\end{align}
where $\int_{\bf k}=\int\frac{{\rm d}^3k}{(2\pi)^3}$.
It can be seen that the result fits the general formula (\ref{mass}). At very high temperatures it reduces to
\be\label{yuk}
m_\phi^2=\frac{y^2+y'^2}6\, T^2   ~~~ (T\gg m_\psi,m_\chi)\,.
\ee
In the opposite regime one has the usual Boltzmann suppression, captured by the asymptotic formula
\be
\int_{\bf k}\frac{n({\bf k})}{E_{\bf k}}\sim \sqrt{\frac{m T^3}{(2\pi)^3}}\,e^{-m/T}\,,
\ee
which is valid for bosons as well as fermions.

It is enough for one among $\psi$ and $\chi$ to decouple from the plasma to have a suppression of the thermal mass (not exponential, though). In the regime $m_\chi\ll T\ll m_\psi$ one has
\be
m_\phi^2=\frac {m_\chi}{ m_\psi} \frac{y'^2-y^2}6\, T^2 \,,
\ee
suppressed by the small ratio $m_\chi/m_\psi$: though $\chi$ is abundantly produced thermally and is available for scattering with $\phi$, the heavy $\psi$ has to be produced virtually in order for the scattering to take place.


\subsection{Gauge interactions}
The scalar $\phi$ can also be charged under some gauge group $G$, abelian or not. The interaction of $\phi$ with the gauge degrees of freedom, and with itself via the gauge coupling, gives another contribution to its thermal mass. The interaction -- mediated by the exchange of a gauge boson -- with other species charged under $G$, instead, gives no contribution at one loop due to an adaptation of Furry's theorem \cite{Baratella:2024sax}.

Using (\ref{mass}) one finds (the covariant derivative is defined as $\partial_\mu-ie A_\mu$)
\be\label{gauge}
m_\phi^2=e^2\int_{\bf k}\frac1{2|{\bf k|}}\bigg(2 n_{\gamma_+}+2n_{\gamma_-}+ n_\phi+n_{\bar\phi}\bigg)=\frac{e^2T^2}{4},
\ee
where the last expression assumes negligible $\phi$ mass, and $\gamma_\pm$ denote photons with respectively positive and negative helicity. If $G$ is non-abelian and $\phi$ carries representation $t^a$, such that the covariant derivative is $\partial_\mu-iet^aA_\mu^a$, one needs to multiply (\ref{gauge}) by $C_\phi$, with $\sum_a t^at^a=C_\phi {\bf 1}$. For example, for $\phi$ in the fundamental of $SU(N)$, $C_\phi=\frac1{2N}(N^2-1)$.

\section{Examples}\label{sec:examples}

\subsection{Scalar decoupling}\label{exscal}

The simplest example we consider is a theory in which $\phi$ couples to a neutral scalar $\varphi$. The relevant physical parameters are captured by the potential
\be\label{Lscalar}
V=-\lambda |\phi|^4+\tfrac12\lambda' \varphi^2|\phi|^2+\tfrac12m_0^2\varphi^2+\lambda'' \varphi^4\,.
\ee
At finite temperature, on top the zero temperature loop corrections (fine tuning issues, which are expected to arise here, are ignored in this discussion) one has to take into account thermal effects. In particular, a thermal mass for $\phi$ is generated, and the mass of $\varphi$ gets a correction that dominates for $T\gg m_0$. In this regime, assuming for simplicity that $\lambda'\gg\lambda,\lambda''$ one has
\begin{align}
m_\phi^2&= \tfrac1{24}\lambda' T^2\,, \\
m_\varphi^2&= m_0^2+\tfrac1{12}\lambda' T^2\,.
\end{align}
Along with $\phi$ and $\varphi$, let us assume that there is a number of other species, not charged under $Q$, which are kept in thermal equilibrium and are either massless or with negligible mass with respect to $m_0$. They have couplings to $\phi$ that are weaker than $\lambda'$ and overall give a thermal mass to $\phi$ that has size $a T^2$, with $a\ll \lambda'$.

When temperature falls below $m_0$, one bosonic degree of freedom decouples from the plasma. More dramatically, the mass to temperature ratio of $\phi$ drops from $\sqrt{\lambda'/24}$ to $\sqrt a$, due to (\ref{drop}).

As a measure of the effect that the decoupling of $\varphi$ has on the balance among temperature and chemical potential, we consider
\be
\rho\equiv\frac{\kappa{(T\ll m_0)}}{\kappa(T\gg m_0)}=\sqrt{\frac{\lambda'}{24 a}}\left(1-\frac{2\pi^2}{45g}\right)\,,
\ee
with $\kappa$ given in (\ref{master}). The new quantity $\rho$ measures if the drop in the thermal mass of $\phi$ is more or less sizeable than the decrease in the number of relativistic degrees of freedom. Unlike $\kappa$, $\rho$ depends only on the specifics of the model and not on the initial conditions.

For the decoupling of $\varphi$ to induce condensation, it is necessary (but not sufficient) that $\rho>1$. If $\rho$ is only slightly larger than 1, condensation is unlikely because one needs the system to be already at the edge of condensation for $T$ above $m_0$ (that is, $\kappa(T\gg m_0)$ must be very close to 1). For larger and larger $\rho$, the range of favourable initial conditions, say on the charge to entropy ratio, gets bigger and bigger. Since the effect is multiplicative, one needs always \emph{some} charge asymmetry for it to operate.
Under the natural assumption that $g\gg1$, if we take $\lambda'\sim1$ then $\rho\sim 2$ for $a=10^{-2}$ and $\rho\sim 6.5$ for $a=10^{-3}$.

While it is natural for the thermal mass to be dominated by the effect of one coupling, it comes as a coincidence that the particle that contributes mostly to $m_\phi$ decouples first from the plasma.


\subsection{Charged fermion multiplet}\label{sec:mult}

As a second illustration of the effect we consider the model of (\ref{Yukawa}) with the symmetric assignment $q_\psi=-q_\chi=\frac12q_\phi$. To enrich the model we also make $\psi$ and $\chi$ multiplets of some non-abelian group $G$. For $T\gg m_\psi,m_\chi$, one has
\be
m_\phi^2=\frac{y^2+y'^2}6\,d_\psi T^2
\ee
with $d_\psi$ the length of the fermion multiplets.

When temperature drops below the mass of fermions, three changes take place. First, $8d_\psi$ fermionic degrees of freedom decouple from the thermal plasma. Second, the thermal mass of $\phi$ drops to $aT^2$, with $a\ll y^2+y'^2$. As a third effect, all of the charge excess of the system gets transferred to $\phi$, while at $T\gg m_\psi,m_\chi$ this was democratically distributed among the various degrees of freedom (we are assuming that no other particle in the plasma besides $\phi,\psi$ and $\chi$ carries $Q$ charge).

Due to the third phenomenon, which was absent in the previous example, the formula for $\rho$ includes another potentially large ratio, coming from correcting (\ref{master}) with (\ref{charge2}). In this model, it is given by
\be
\rho=\sqrt{\frac{(y^2+y'^2)d_\psi}{6a}}\bigg(1-\frac{14\pi^2d_\psi}{45g}\bigg)(1+d_\psi)\,,
\ee
where the last factor is ${q_\phi^{-2}}{\cal Q}^2(T\gg m_\psi,m_\chi)$, with ${\cal Q}^2$ as defined in (\ref{charge2}).
Even with a sizeable decrease in the number of relativistic degrees of freedom, say by a factor of 2, a large multiplet size induces a much more drastic effect than in the example of Section~\ref{exscal}. For $y,y'\sim1$, with $a=10^{-2}$ one has $\rho\sim 39$ and $\rho\sim 100$ for respectively $d_\psi=5,10$. The effect is sizeable also for a less dramatic decrease in the thermal mass. With $a=10^{-1}$, $\rho\sim 12$ and $\rho\sim 32$ for respectively $d_\psi=5,10$. As a matter of fact, charge redistribution is a distinct effect that takes place whether or not the thermal mass of $\phi$ changes significantly.

In general, both effects are present and the first gives a boost to the second, or vice versa.

\subsection{Dynamically generated exit scale}\label{sec:dyn}

In the present section we explore the possibility that particle decoupling from the plasma is due to a dynamically generated mass scale.

Consider a gauge theory, charged under local $SU(3)\times SU(2)$, whose particle content is that of one generation of Standard Model fermions and one Higgs doublet $\phi$. Using the familiar Standard Model nomenclature, let us call the gauge group multiplets $q(3,2)$, $u^c(\bar 3,1)$, $d^c(\bar 3,1)$, $\ell(1,2)$, $e^c(1,1)$ and $\phi(1,2)$.

Let us assume that the only Yukawa term that has non-negligible coupling to $\phi$ is
\be
{\cal L}_Y=y_e \phi^*\ell e^c\,.
\ee
In this scenario there are several conserved $U(1)$ charges, that can be chosen to be the `baryon numbers' $B_q,B_u,B_d$, which count the number of respectively $q,u^c,d^c$ (minus their antiparticles), the lepton number $L$, which counts the number of $\ell$ minus the number of $e^c$, and a charge $Q$ under which $q_\phi=1$ and $q_\ell=q_{e^c}=\frac12$. Let us assume that $Q\neq 0$, while $B_q,B_u,B_d,L=0$.


The interesting scale for inducing condensation is the analogue of $\Lambda_{\rm QCD}$ in this toy example, which is assumed to have roughly the same features of the real-world QCD (with two flavours of quark, the one-loop beta function of $\alpha_s$ is $\mu\frac{{\rm d}}{{\rm d}\mu}\alpha_s=-\frac{\alpha_s^2}{2\pi}(11-\frac43)$, making $\alpha_s$ grow in the infrared). For $T\sim\Lambda_{\rm QCD}$ confinement is assumed to take place with generation of a chiral condensate \cite{Ciambriello:2022wmh}. 

The formation of a quark condensate was crucial in the context of \cite{Witten:1980ez} to give a linear term to the potential of $\phi$ via its Yukawa coupling to the quarks. In the hypothesis that such a direct coupling to the condensate is absent or very small, a more efficient exit mechanism is provided, in the right conditions, by the supersaturation of $Q$. Let us see how this arises.

The chiral condensate has the same quantum number of the Higgs and it breaks $SU(2)_L$ completely, giving a perturbative mass to the three associated gauge bosons $W^i_\mu$ via the Higgs mechanism. The scalar degrees of freedom that are eaten by the $SU(2)_L$ gauge bosons are the three pions $\pi^i$, which can be embedded in $U=e^{i\pi^i\sigma^i/f_\pi}$. The canonically normalised gauge covariant kinetic term of $U$ reads $\frac{1}4f_\pi^2{\rm tr} D_\mu U^\dagger D_\mu U$, with $D_\mu U=\partial_\mu U-ig_2W_\mu^i\frac{\sigma^i}{2}U$. Taking $\langle U\rangle={\bf 1}$, the kinetic term produces a mass term for the $W$ bosons,
\be
{\cal L}_{m_W}=\frac12\bigg(\frac{g_2f_\pi}{2}\bigg)^2\delta_{ab} W_\mu^aW_\mu^b\,.
\ee
What is important for us is that QCD confinement induces, indirectly via the chiral condensate, a mass to the $W$ bosons. Therefore, for $T\ll m_W=\frac12g_2f_\pi$, the only active species in the plasma are $\ell,e^c$ and $H$.

The drop in the number of active degrees of freedom is considerable, and is quantified by
\be\label{dyn_g}
\frac {g'}g=\frac{37}{165}\simeq 0.22\,.
\ee
If $g_2\gg y_e$, the thermal mass to temperature ratio also drops significantly, giving a contribution to $\rho$ of order
\be
\frac{\sqrt{a}}{\sqrt {a'}}\sim \frac{g_2}{y_e}\gg 1\,,
\ee 
so that the effect of (\ref{dyn_g}) can be overcome. Let us see how this comes about. For $T\gg m_W$, the main contribution to $m_\phi$ is given by its interaction with $W$ particles in the plasma and self interactions mediated by the exchange of $W$ bosons. Using (\ref{gauge}) and that $C_\phi=\frac 34$ for a fundamental of $SU(2)$, one finds $m_\phi^2(T\gg m_W)=\frac3{16}g_2^2T^2$. For $T\ll m_W$ the contribution to $m_\phi$ coming from interactions with gauge bosons is Boltzmann-suppressed, and the one coming from self-interactions mediated by $W$ is expected to be suppressed by the ratio $T^2/m_W^2\ll 1$ because $W$ has to be produced with high virtuality. Therefore the main contribution to the thermal mass of $\phi$ comes from interactions with $\ell$ and $e^c$ mediated by the Yukawa coupling $y_e$. Using (\ref{yuk}) with $y,y'\to \frac{y_e}2$ one finds $m_\phi^2(T\ll m_W)=\frac1{12}y_e^2T^2$.

In this example no $Q$-charged particle Boltzmann-decouples from the plasma, so one gets simply
\be
\rho=\frac{37g_2}{110y_e}\,.
\ee

\section{Conclusions}\label{sec:conclusions}

A long period of supercooling has most of the qualitative features of Inflation: it is a period of inflationary expansion that lasts a number $N_e\sim \ln T_c/T_{\rm exit}$ of $e$-folds, where $T_{\rm exit}\sim m_0,m_{\psi,\chi}$ and $\Lambda_{\rm QCD}$ in the examples of \ref{exscal}, \ref{sec:mult} and 
\ref{sec:dyn} respectively. 
When the condition for condensation is reached, the system starts relaxing to the true minimum and all latent heat gets released, increasing the entropy by a factor of roughly $(T_c/T_{\rm exit})^3$. At the end of the reheating period $Q\sim (T_{\rm exit}/T_c)^3S\ll S$ (the onset of condensation requires, very roughly, that $Q\sim S$). A similar dilution takes place for any other abundance or defect produced before the supercooling epoch.

The transition to the stable vacuum is induced by tachionic instability \cite{Felder:2001kt}.
The out-of-equilibrium dynamics that follows the sudden destabilisation is potentially rich in implications and deserves further investigation. In the hypothesis of this work, $\phi$ is $U(1)$-charged and the transition to the true vacuum is expected to give rise to a string network, similarly as in the Peccei-Quinn phase transition \cite{Gorghetto:2023vqu}. Adding a small $U(1)$ violating term that allows the network to eventually annihilate is not expected to spoil the mechanism described in this work.

The idea that a strong chemical potential can induce the destabilisation of a metastable vacuum has been pursued in \cite{Balkin:2021zfd}; the authors considered large and dense systems such as stars as catalysers for the formation of true-vacuum bubbles. Here, instead, the presence of a sizeable charge concentration characterises the whole universe at some time in its evolution.

In the context of Inflation, the authors of \cite{Bodas:2020yho} have shown how the large kinetic energy of the inflaton can mimic a chemical potential $\mu\gg T_{\rm dS}$ which is able to overcome Boltzmann suppression of particles with mass $m> T_{\rm dS}$; the condition $\mu<m$ is invoked there to avoid tachionic instabilities. It could be interesting to study an implementation of the mechanism presented here where the destabilising chemical potential comes from the coupling of $\phi$ with another scalar $\varphi$ that has $\dot\varphi\neq 0$.

A period of supercooled inflation has phenomenological consequences even if it happens in a sector that is only gravitationally coupled to the Standard Model; see for instance \cite{Garani:2021zrr} for a discussion on the cosmology of dark sectors. As a model-building tool, supersaturation allows a metastable dark sector to persist until it reaches a critical threshold and decays.

The dynamical generation of a confinement scale -- what happens in QCD -- is a generic mechanism in gauge theories that allows to accommodate large hierarchies. In the toy example of \ref{sec:dyn}, a larger and larger hierarchy between the critical temperature $T_c$ and the exit temperature $T_{\rm exit}\sim\Lambda_{\rm QCD}$ implies a longer and longer period of supercooling with inflationary expansion. It would be interesting to study what ingredients are needed, other than $N_e=\ln T_c/T_{\rm exit}\gtrsim 50$, to embed the supersaturation mechanism in a realistic model for Inflation.


\

{\bf \textsc{Acknowledgments.}}
I thank Jernej Kamenik for discussions and comments on the manuscript, and Takeshi Kobayashi, Mehrdad Mirbabayi, Lorenzo Di Pietro, Alex Pomarol and Nicklas Ramberg for discussions. I am grateful to Reuven Balkin, Javi Serra and Stefan Stelzl for sharing stimulating thoughts on the role of chemical potential in cosmology. This work is supported by the Slovenian Research Agency (research core funding No. P1-0035 and J1-4389).

\bibliographystyle{apsrev4-2}
\bibliography{references}

\end{document}